\documentstyle[twocolumn,prc,aps,epsfig]{revtex}
\input epsf

\def\subs{\underline{S}}
\def\subu{\underline{U}}
\def\subd{\underline{D}}
\def\subq{\underline{q}}

\newcommand{\be}{\begin{eqnarray}}
\newcommand{\ee}{\end{eqnarray}}

 \newcommand{\gsim}{\mathrel{\hbox{\rlap{\lower.55ex \hbox {$\sim$}}
                   \kern-.3em \raise.4ex \hbox{$>$}}}}
\newcommand{\lsim}{\mathrel{\hbox{\rlap{\lower.55ex \hbox {$\sim$}}
                   \kern-.3em \raise.4ex \hbox{$<$}}}}
\newcommand{\la}{\langle}
\newcommand{\ra}{\rangle}

\def\roughly#1{\mathrel{\raise.3ex\hbox{$#1$\kern-.75em%
\lower1ex\hbox{$\sim$}}}}
\def\lsim{\roughly<}
\def\gsim{\roughly>}

\def\la{{\Big<}}
\def\ra{{\Big>}}

\begin{document}

\twocolumn[\hsize\textwidth\columnwidth\hsize\csname @twocolumnfalse\endcsname

\title{A schematic model for pentaquarks based on diquarks}

%\thanks{xxx}

\author{Edward Shuryak and Ismail Zahed}

%\vspace{0.4cm}
\address{
Department of Physics and Astronomy,\\ State
University of New York at Stony Brook, New York 11794, USA }

\maketitle
\begin{abstract}
\noindent 
QCD instantons are known to produce deeply bound diquarks which
 may be used as building blocks in the
formation of multiquark states, in particular pentaquarks and dibaryons.
We suggest a simple model in which the
lowest scalar diquark (and possibly the tensor one)
can be treated as an independent ``body'', with 
the same color  and (approximately) the mass as a
constituent (anti)quark. In this model a new symmetry between
states with the same number of ``bodies'' but different number of
quarks appear, in particular the 3-``body'' pentaquarks can be 
naturally related to decuplet baryons. We estimate both the 
masses and widths of such states, and then discuss 
the limitations of this model.
\end{abstract}
\vspace{0.1in}
]
%\begin{narrowtext}
%\newpage

\vskip 1.5cm

{\bf Introduction.\,\,\,}
The possibility of a low lying $\bar q q^4$ states in the P-wave (e.g. $K^+n$) 
channel fitting in the anti-decuplet flavor representation of the quark
model was  advocated long ago by Golowitch~\cite{GOLO}, along with the 
non-strange excited baryon $N(1710)$. 
A decade ago, when the $SU(3)$ version of the Skyrme 
model was refined, it was found to predict an
{\em antidecuplet} $\bar{10}$ of baryons  above 
the conventional octet and decuplet.
It was not taken seriously till relatively recent
works~\cite{SKYRME} which predicted among others
a resonance in  $K^+n$ with a mass of 1540 MeV.

In remarkable agreement with this prediction,
several recent experiments have reported an
exotic baryon $\Theta^+(1540)$ with a small 
(and so far unmeasured) width~\cite{exp}.
The issue of its consistence with earlier $Kd$ data is discussed
in~\cite{Nussinov} and also~\cite{KARLINER}.  The observed angular 
distribution suggests a likely spin $1/2$ state, with so far unknown parity. 
Its minimal quark content is a pentaquark, i.e. $(ud)^2\bar{s}$.  The 
antidecuplet flavor assignment was further strengthened by an observation by
the NA49 collaboration~\cite{NA49-03} of a family  of exotic $\Xi$ baryons, 
with a mass of 1.86 GeV and width smaller than the experimental resolution
of 18 MeV.

The theoretical advantage of the Skyrme model is that it allows to
reduce a  complex multiquark problem into a 
 single-body problem, with one pseudoscalar meson moving in
a fixed classical background. However the price for such reduction,
based on the ``large $N_c$ ideology'' maybe prohibitive given the
large degeneracies implied. The $1/N_c$ description implies a small
width, that is difficult to assess quantitatively given the subtleties
related to these corrections~\cite{COHEN}.
%one may not only question its accuracy but the overall picture
%as well. Indeed, pentaquarks made of the usual baryons and
%pseudoscalar
%mesons have no reason to be as narrow as the observed states.

More traditional ``shell model ideology'' (e.g. the MIT bag model or
nonrelativistic constituent quark models) tends to put as many quarks
as possible into the lowest shell, and thus predict $negative$ parity, $P=-1$
for the lowest state\footnote{The lattice studies
by Csikor \emph{et al.}~\cite{Csikor:2003ng} and Sasaki~\cite{Sasaki}, 
indeed claim a signal for $P=-1$  pentaquarks with a comparable mass. More
and better data
are however needed to reach firm conclusions on the matter. }.

 The shell model works well 
for nuclei; in this case the  pairing effects are small and treated
perturbatively using the shell model states. However we think
the order should be reversed for hadrons, and pairing into diquarks 
be treated first. One argument for that is that
the many flavor-symmetric  exotic states  possible in a shell 
model have never been  seen. Even the most symmetric ``magic'' configuration,
the dibaryon $H=u^2d^2 s^2$, an analogue of the alpha particle, appears
to be not deeply bound, as it was never found in multiple decicated searches. 

\begin{figure}
 \includegraphics[width=8.cm]{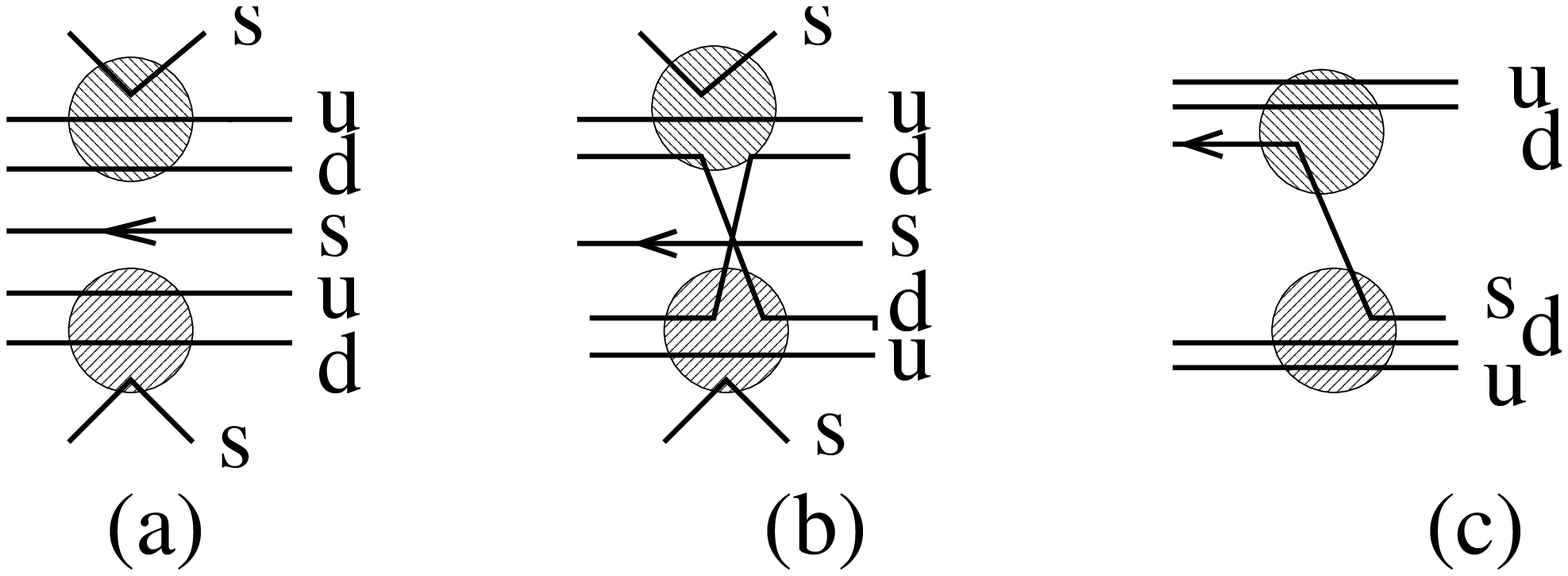}%
 \caption{\label{fig_paths}
Some second-order instanton-induced interactions
of  5 quarks propagating in time through the
(Euclidean) QCD vacuum. The shaded circles indicate instantons and 
antiinstantons. The quarks are avoiding
quarks of the same flavor and 3-body force is repulsive, so (a)
is the diagram generating two independent diquarks.
 The instantons have to pick up pairs from the vacuum condensate
$<\bar s s>$ to get it attractive. The diagram (b)
 with a light quark exchange generates a repulsive core, while
the diagram (c) leads to  diquark attraction.
}
 \end{figure}

As we will argue in this letter, the picture most
consistent with the current new findings are those
developed in a ``small $N_c$ ideology'', in which the key element
are the {\em instanton-induced}~\footnote{Although scalar diquarks are
also attracted by single-gluon exchange forces, the latters 
do not lead to the structure we discuss as they are flavor blind.}
 diquarks \cite{SHU,RILM:SS98}.
Due to the Pauli principle at the level of instanton zero modes,
two quarks of the same flavor cannot interact with the same instanton.
The propagation of 5 quarks through the QCD vacuum generates many
interactions involving 't Hooft interaction, some second order ones are
depicted in Fig.~\ref{fig_paths}. The latter illustrates  the strong preference
for multiquark states to be in the lowest possible flavor representation,
avoiding many other possible exotic states, both 
in the meson and baryon sectors. As we will argue,
even these newly discovered  states, although truly exotic, still
are in a way analogous to the decuplet baryons. Their small decay
widths is a consequence of a {\em  different internal
structure}, with small overlap with all the decay channels. 
 
For a  review on the instanton vacuum models one can consult~\cite{RILM:SS98}.
The main approximations  are: {\bf i.} a reduction of the
gauge configurations to the subset of instantons and antiinstantons;
{\bf ii.} a focus on only the fermionic states that are a superposition
of their zero modes. When
the baryonic (3-quark) correlators have been first calculated 
in it \cite{RILM:BaryCor} a decade ago
(and soon confirmed by  
lattice measurements \cite{RILM:BaryLat})  a marked difference between
the nucleon (octet) and $\Delta$ (decuplet) correlators has been
noted. Roughly speaking, a nucleon was found to be
made of a quark and a   
 very deeply bound {\em scalar-isoscalar diquark}, absent in
the decuplet.  As it was found to have a
surprisingly small mass comparable to 
 the constituent quark mass 
(to be denoted below as $\Sigma$), it significantly simplifies 
the model to be discussed below.

The general theoretical reason
 for the lightness of the scalar-isoscalar diquark
state (see e.g.\cite{DIQ:BoseCond}) follows from the special Pauli-Gursey 
symmetry of 2-color QCD. In this theory (the ``small 
$N_c$ limit'' of QCD) the scalar diquarks are actually 
{\it massless Goldstone bosons}. For general $N_c$, the 
instanton (gluon-exchange) in $qq$ is $1/(N_c-1)$ down relative to
$\bar{q}q$. So the real world with $N_c=3$ is half-way between
$N_c=2$ with a relative weight of 1, and $N_c=\infty$ with 
relative weight 0. Loosely speaking, the scalar-isoscalar
diquarks are {\it half Goldstone bosons} with a binding energy of about
half of the mass, or about one constituent quark mass.

Diquarks in the
context of Nambu-Jona-Lasinio models were investigated e.g. 
in~\cite{DI}, which also emphasized the occurrence of a light 
scalar-isoscalar bound state.
Diquark correlations have been a driving idea behind a view of dense
baryonic matter as a very strong color superconductor
\cite{DIQ:BoseCond,ARW}. If one views the nucleon as a quark plus a Cooper
pair, such a view of dense matter is indeed very natural. 
In hadronic spectroscopy the nonet of scalar mesons below 1 GeV
is belived to be made of diquark-antidiquark states.

In such a context it is even more natural to see the pentaquarks
as an antiquark plus $two$ Cooper pairs.
Jaffe and Wilczek (JW)~\cite{JW} (see also
 Nussinov~\cite{Nussinov}) have already suggested
to view the $\Theta^+(1540)$ as an object made of 2 diquarks
$(ud)(ud)\bar{s}$, where $(ud)$ is a  $scalar$ isoscalar
diquark in relative P-wave. This model leads to an ${\bf 8}_f\oplus 
{\bf{\overline{10}}}_f$ flavor representation for the pentaquark states. They
also argued that the long-known Roper 1440, may also be a 
$(ud)^2\overline{d}$ pentaquark state belonging to an octet. 
In a more recent paper \cite{JW2} they have added further considerations
following from the Na49 cascade data: the most important one is that they
seem to provide experimental indications
on the existence of the pentaquark octet, together with
 $\bf{\overline{10}}$.

In this letter we develop these ideas a bit further,
suggesting a schematic model which has enough symmetries
to allow estimates of the pentaquark  masses by relating them
to those of decuplet baryons. Our input are  the values 
for the ``diquark masses'', calculated in
 the random instanton liquid model (RILM)~\footnote{Those exist as physical hadrons only in
  $N_c=2$ QCD. However, since  the instanton liquid
  model does not confine,  there are diquark states for any $N_c$.}.

\vskip .5cm

{\bf $\overline{3}_c$ diquarks\,\,.} All diquarks to be discussed 
below are {\it anti-triplets} in color (both instanton and gluon
interactions are repulsive in the sixteth) with generic spin-flavor 
assignments as follows

\be
(q{\bf \Gamma}q)^a=\epsilon_{abc}\,q_b^T{\bf C}{\bf \Gamma} \,q_c\,\,,
\label{1}
\ee
where ${\bf C}$ is the charge conjugation matrix, and ${\bf \Gamma}$
include the pertinent Dirac and flavor matrices. 
Diquarks with all possible Dirac matrices ${\Gamma}$
in $ q^T C{\Gamma} q $ have been studied in RILM~\cite{RILM:BaryCor}.
The pseudoscalar channel with $\Gamma=1$ was found to be very strongly
repulsive, the vector and axial vector channels are weakly
repulsive, with a mass of the order of 950 MeV,
above twice the constituent quark mass of the model, 
$2\Sigma=840\, {\rm MeV}$. The only two channels with
attraction and  significant binding are: 
{\bf i.\,} the {\it  scalar} with $m_S\approx\Sigma$ and 
$\Gamma =\gamma_5$; {\bf ii.\,} the {\it  tensor} with
$m_T\approx 570\, {\rm MeV}$ and  $\Gamma=\sigma_{\mu\nu}$ (denoted below
by a subscript $T$)~\footnote{The longitudinal vector diquark channel with
$\Gamma=\gamma_\mu\gamma_5$ mixes with the scalar $\Gamma=\gamma_5$ in
the P-wave. This point is relevant to the lattice studies
discussed in~\cite{Sasaki}}. 
The scalar is odd under spin exchange while the tensor is even under
spin exchange. Fermi statistics forces their flavor to be different.
The scalar is flavor antisymmetric $\bar 3$ while  the tensor
is  flavor symmetric $6$. 

In the model to be discussed below, we will discuss all possible pentaquark
multiplets which can be made using these ingredients. For scalar diquarks
we will introduce the following shorthand notation in $SU(3)_f$

\be
\subs=(u^T C\gamma_5d); \,\,\,\subu=(s^T C\gamma_5d); \, \,\,\subd=(u^T
C\gamma_5s)
\ee
 
\vskip 0.5cm

\begin{figure}
 \includegraphics[width=8.cm]{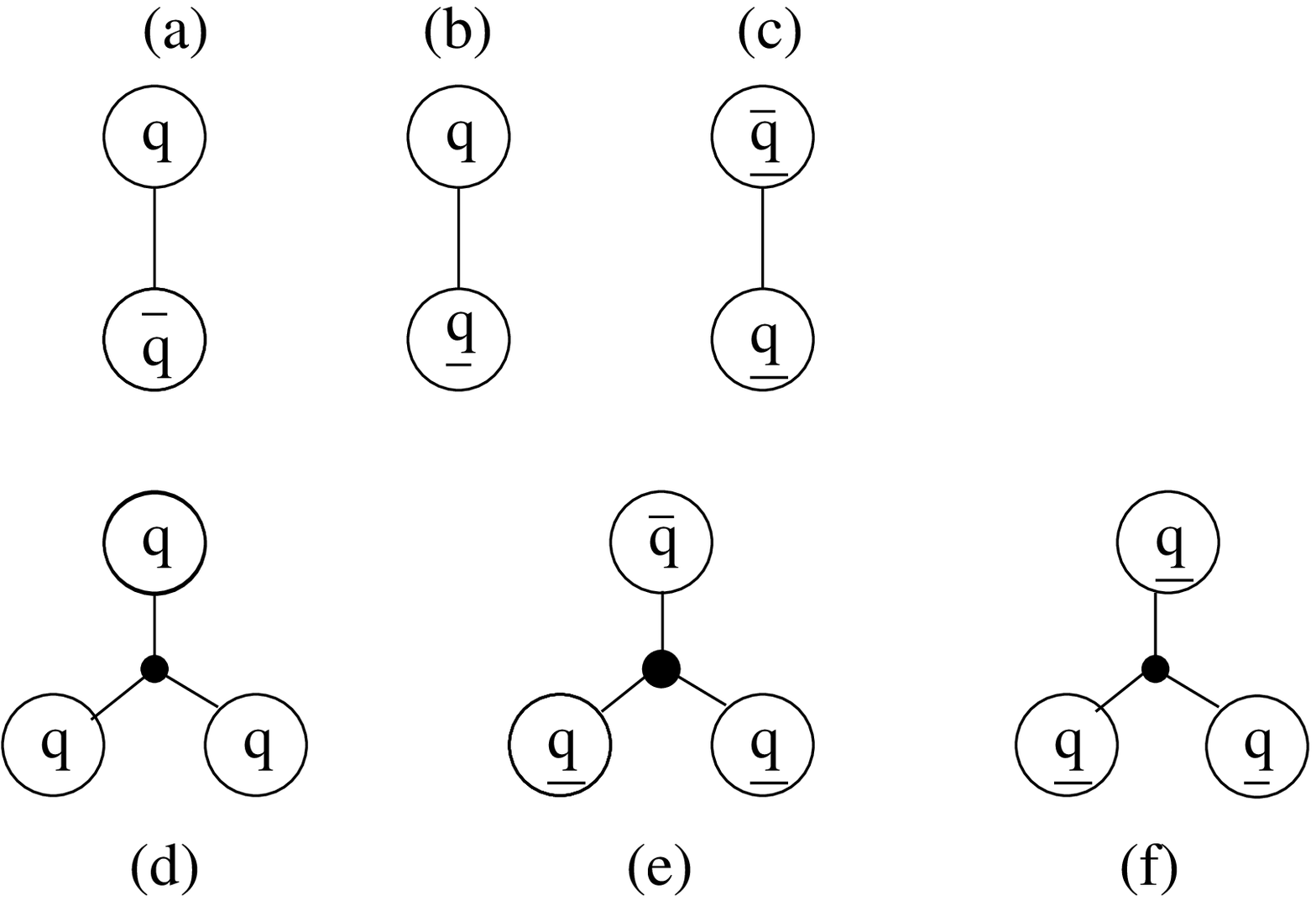}%
 \caption{\label{23bodies}
Schematic structure of (a)  ordinary mesons, (b) quark-diquark or
octet baryons, (c) diquark-antidiquark states or tetraquarks, (d) decuplet baryons, (e)
pentaquarks and (f) dibaryons. 
}
 \end{figure}

{\bf Model\,\,.} to be discussed  treats 
 diquarks  on equal footing with constituent quarks.
Because of their similar mass and quantum numbers,
certain approximate symmetries appear between states 
with the same numbers of ``bodies''. This simple idea 
 is depicted pictorially in Fig.~\ref{23bodies}.
The $\bar q q $ mesons (a) are
a well known example of the 2-body objects,
as well as the quark-diquark states (b) (the octet baryons $\subq q$).
The diquark-antidiquark states (c) are in this model
the 2-body objects.  In $zeroth$ order,
the usual non-strange mesons (like $\rho,\omega$),
   the octet baryons (like the nucleon), and the 4-quark  mesons
(like $a_0(980)$)\footnote{For recent study of
     these states in the instanton model
 see \cite{Schaefer:2003nu}.})
are degenerate, with a mass $M\approx2\Sigma= 840\, MeV$.
To  $first$ order, which includes color-related interactions, the  
one-gluon-exchange Coulomb and confinement, the degeneracy should still hold,
as the color charges and the masses of quarks and diquarks are the same.
 Only in $second$ order, when the spin-spin and other residual
 forces are included, they split. There is no spin-spin interaction for the nucleon
(the scalar diquark has no spin), while for
 the $\rho$ it is either repulsive (if it is due to one gluon exchange) or
zero (if it is due to the instanton-induced forces~\cite{spinforces}).
Note that this new  symmetry 
between $N$, $\rho$ and $a_0(980)$ is actually rather accurate, better than
the old SU(6) symmetry, stating (in zeroth order)
that  $M_N\approx M_\Delta$.  

{\bf  Pentaquarks\,\,\,} in such a model are treated as
3-body objects, with two correlated  diquarks plus 
an antiquark, and thus  there are
 simple relations between
masses of various ``3-body objects'' depicted in  Fig.~\ref{23bodies}
(d-f) with  the ``3-body'' (octet/decuplet) baryons.

From the color point of view,  all 3-body
states  involve the same $\epsilon_{abc}$ wave function, just
like the ordinary color singlet baryons. From the flavor point of
view the situation is different. For pentaquarks made of two
scalar diquarks the flavor representations are
$\bar 3\otimes \bar 3 \otimes \bar 3=1\oplus 8\oplus 8\oplus \bar 10$.
Using the   notations we introduced above, and changing from bar to underline
where needed,  one can readily see how the pentaquarks observed fit onto
an antidecuplet, $\Theta^+(1540)=(ud)(ud)\bar{s}=\subs \subs \bar{s}$
is an analogue of anti-$\Omega$, and is thus
the top of the antidecuplet (the conjugate of the decuplet). 
New exotic  $\Xi(1860)$ are $\subu \subu \bar u$ and    $\subd \subd
\bar d$, providing the two remaining corners of the triangle. They are
the analogue of anti-$\Delta$.  The remaining 7
members can mix with one octet, as discussed by Jaffe and Wilczek,
making together 18 states in  flavor representations
$({\bf 8}\oplus {\bf {\overline 10}})$.
For ordinary 3 quarks there is the overall Fermi statistics
which ties together flavor and spin-space symmetry and works against
the remaining $1\oplus 8$. There is no such argument for
pentaquarks. So how are the additional flavor states 
$1\oplus 8$ excluded for pentaquarks?
 
For diquark-diquark-antiquark all there is left is Bose statistics 
for identical  scalars, demanding total symmetry over their interchange, 
while the color wave function is antisymmetric. So
the only solution~\cite{JW,Nussinov} is to make the spatial  wave
function antisymmetric by putting one of the diquark into the P-wave
state. It means that such pentaquarks should be 
degenerate with the excited P-wave decuplet baryons.

\be 
M_\Theta= 2\Sigma+\Sigma_s+\delta M_{L=1}+V_{residual}
\ee
where the first 2 terms are masses of the diquarks and strange quark,
plus an extra  contribution for the P-wave, plus whatever {\it residual}
interaction there might be.

It is straightforward  to assess $\delta M_{L=1}$ by analogy with
the  P-wave baryon excitations. Indeed,
the diquark mass is about the constituent quark mass, and the
confining potential is also the same. For example, 
following the well known paper by Isgur and Karl~\cite{IK_78} one can
simply use an oscillator potential, in which the separation of the center of mass
motion from the internal motion is relatively simple.
Introducing three standard Jacobi coordinates,
one finds that the difference between P-wave and
S-wave state is $\delta M_{L=1}=\hbar \omega_\lambda\approx 480 \, MeV$.
Very similar values were obtained using more modern constituent quark models, e.g.
a semi-relativistic model with a
linear potential by the Graz group\cite{Glozman:1997xh}\footnote{ The 
difference with Isgur and Karl is in the nature of the spin-spin forces which are not
important  for scalar (spinless) diquarks.}, 
so we consider our assessment justified.

Ignoring for the time being all residual interactions,
one may estimate the pentaquark mass to be that of a decuplet baryon
with a single $s$ plus the P-wave penalty, i.e.
\be 
m_\Theta \approx m_\Sigma^*(3/2)+\delta M_{L=1}\approx 1400+480=1880 \,
{\rm MeV} \,\,,
\ee
which is  well above the  observed mass of 1540 MeV.

However, using one scalar and one {\it tensor} diquark one can do
without  the P-wave penalty, and the schematic mass estimate now reads
\be  
m_\Theta \approx m_\Sigma^*(3/2)+\delta M_{T} \approx
 1400+150=1550\, {\rm MeV}\,\,, 
\ee
which is much closer to the experimental value.

The  newly observed $\Xi(1860)$ pentaquarks contains diquarks with a
strange quark, that is 
$us,ds$. Their masses have not been yet directly calculated, but
a general experience with spin-dependent forces\cite{spinforces} suggests a
reduction of binding by about a factor 0.6 as compared to the $ud$
case. This suggests a total loss of binding of about 200-240 MeV,  
which together with a strange quark mass itself (two s quarks instead
of a single $\bar s$) readily explains the 320 MeV mass difference
between  $\Xi(1860)$ and $\Theta^+(1540)$  pentaquarks.

Since the tensor diquark has the opposite parity, both possibilities
correspond to the same global parity $P=+1$. Also common to both
schemes is the fact that the total spin of 4 quarks is 1, so
adding the spin of the $\bar s$ can lead not only to $s= 1/2^+$ but also to
$s=3/2^+$ states (which are not yet observed). 

So, we conclude that if we only look at the {\it masses}, it appears that
it is better to substitute one diquark by its tensor variant, rather
than enforce the P-wave. 
 However such an alternative scheme provides a different set of
 flavor representations as we now show. Indeed,
 $\bar 3\otimes 6 \otimes \bar 3=1\oplus 8\oplus 8\oplus 10\oplus 27$.
The largest representation 27 has particles with
quantum numbers of $\Theta^+$ and $\Xi(1860)$, and even more exotic
triplets such as $\Omega$-like $sssq\bar q$ states.
The cascades have isospin 3/2, as observed. However $\Theta^+$
is a part of an isotriplet,
with $\Theta^{++}$ and $\Theta^0$ partners. The former can decay into
$pK^+$, a quite visible mode, in which no resonance close to 1540 was seen.
Since the widths are unknown at this point, it is perheps premature
to conclude that they do not exist.  However if the occurence of
the decay mode $\Theta^{++}\rightarrow pK^+$ is definitively
ruled out, the observed multiplet of exotics cannot be the 27.

%%%%%%%%%%%%%%%%%%%%%%%%%%%%%%%%%%%%

The Roper resonance belongs to octet
 with the quark content $\subs \subs \bar{d}$.
 In the JW model with the P-wave, 
its mass  would be estimated as
\be 
m_{Roper}=&& 3\Sigma+\delta M_{L=1}\nonumber\\
\approx&&  m_\Delta+\delta M_{L=1}=1260+480=
1740\,\,{\rm MeV}\,\,,
\ee
while in a variant with the tensor diquark 
 it is only
\be 
m_{Roper}= &&3\Sigma+ \delta M_{T}\nonumber\\
\approx &&m_\Delta+ \delta
M_{T}\approx 1260+150=
 1410 \,{\rm MeV}\,\,,
\ee
which once again gets us closer to the experimental
 value. However this corresponds to  27 flavor
 representation, where its isospin is 3/2. In the lower 8
 flavor representation with isospin 1/2, 
it will include $\bar s s$ and be too heavy again.

\vskip 0.5cm

{\bf Widths and Goldberger-Treiman Relations\,\,.}
Small widths are not the consequence of the centrifugal barrier,
as the P-wave is not really producing sufficiently small factors.
As we already mentioned, a general argument for 
 small pentaquark widths is  small overlap between the internal and
external ($KN$) wave functions. In this section we make this
relation more explicit.

The decay widths including Goldstone
bosons are determined by  general
properties of their chiral interaction, and expressions can be
somewhat simplified.
The strong decay of the pentaquark $P(\frac 12^+)\rightarrow \pi N
(\frac 12^+)$  is conditioned
by a generalized Goldberger-Treiman relation.  The one-pion reduced
axial vector current has a transition matrix

\be &&\la P(p_2)|{\bf j}_{A\mu}^a (0) |N(p_1)\ra
\nonumber\\
&&=\bar{P}(p_2)\left(\gamma_5\gamma_\mu {\bf G} (t)+ (p_2-p_1)_\mu\,\bar{\bf
H} (t)\right)\frac {\tau^a}2\,N(p_1)
\label{op1}
\ee
with ${\bf j}^a_{A\mu}$ partially conserved~\cite{hide},

\be 
{\partial}^\mu{\bf j}_{A\mu}^a (x) = f_{\pi}\,\left(\Box
+m_\pi^2\right)\,\pi^a(x)\,\,. 
\label{op2} 
\ee 
The first form factor in (\ref{op1}) is one-pion reduced with
${\bf G} (0)={\bf g}_{PN}$ the  ``axial overlap'' charge.
If its value be close to the  axial charge of the nucleon, it would
mean that pentaquark is nothing but a PN system. However, as we
will see, the data demand it to be significantly smaller.

%E I have not changed below this till conclusions

Inserting (\ref{op2}) into (\ref{op1}) gives

\be 
&&\la P(p_2)|\pi^a(0) |N(p_1)\ra  =\frac 1{f_\pi}\frac
{1}{m_\pi^2-t}\nonumber\\&&\times
\bar{P}(p_2) \left((m_P+m_N)\,{\bf G}(t) + t\,\bar{\bf H} (t)\right)\frac {\tau^a}2 N(p_1)\,\,. 
\label{op3}
\ee 
By definition, the pseudoscalar $\pi$-$PN$ coupling is

\be 
&&\la P(p_2)|\pi^a(0) |N(p_1)\ra\nonumber\\&& ={\bf g}_{\pi PN} (t)\,
\frac 1{m_\pi^2-t} \bar{P}(p_2)\gamma_5\tau^a N(p_1)\,\,,
\label{op4} 
\ee 
which corresponds to

\be
{\bf g}_{\pi PN}\,\pi^a\,\left(\bar{P}\tau^a\,N +{\rm
h.c.}\right)\,\,.
\nonumber
\ee
A comparison of (\ref{op4}) to (\ref{op3}) gives at the pion
pole $t\approx m_\pi^2$

\be
f_{\pi}\,{\bf g}_{\pi PN} (m_\pi^2) +\sigma_{\pi PN} (m_\pi^2)= 
\frac {m_P+m_N}2\,{\bf g}_{PN} (m_\pi^2)
\label{op5}
\ee
which is the general form of the
Goldberger-Treiman relation for the transition amplitude
$P\rightarrow N\pi$. The overlap sigma-term is proportional to 
$m_\pi^2/\Lambda$, which is typically 40 MeV in the pion-nucleon
system.
% (for three flavors we are considering
%the F only sigma term).
%For pions $f_\pi\approx 93$ MeV 
%and $m_\pi^2/\Lambda\approx 40$ MeV, while for kaons 
%$f_K\approx 115$ MeV and $m_K^2/\Lambda\approx 500$ MeV. 
%For the pentaquark states ${\bf g}_{PN}\approx 1$, so
%that the strong couplings from (\ref{op5}) are
%
%\be
%{\bf g}_{\pi RN}&&\approx 12.15\nonumber\\
%{\bf g}_{K \theta N}&&\approx 6.5
%\ee
%Similar couplings for the other members of the ${\bf 8}\oplus {\bf 10}$
%plet follows from (\ref{op5}) for an axial transition charge of about 1.

The generic form of the decay width $P\rightarrow \pi N$ is given by

\be
\Gamma_{P\rightarrow \pi N}=\frac{{\bf g}_{\pi PN}^2}{4\pi}
\frac {q_P}{M_P}\,\left(\sqrt{q_P^2+m_N^2}-m_N\right)
\label{dw1}
\ee
where $q_P$ is the meson momentum in the rest frame of the $P$ state,

\be
M_P=\sqrt{q_P^2+m_N^2}+\sqrt{q_P^2+m_\pi^2}\,\,.
\label{dw2}
\ee

The recently observed $\Xi(1860)$ can be used in conjunction
with (\ref{op5}) to bound the transition axial-overlap ${\bf g}_{PN}$ 
and the coupling ${\bf g}_{\pi PN}$ in the antidecuplet, thereby
allowing a prediction for the width of the $\Theta(1540)$
through (\ref{dw1}).
Indeed, if we assign a conservative decay width of about 20 MeV to 
$\Xi^{--}\rightarrow \Xi^-\pi^-$ in light of the bound of 18 MeV
reported by~\cite{NA49-03}, then (\ref{op5}) suggests 
${\bf g}_{\Xi \Xi} \approx 0.25$ and ${\bf g}_{\pi\Xi\Xi}\approx 3.75$
for $\sigma_{\pi\Xi}\approx 40$ MeV. Similar arguments yield 
${\bf g}_{\Xi \Sigma} \approx 0.25$ and ${\bf g}_{K\Xi\Sigma}\approx
2.97$, thus an estimated partial  width of $6.60$ MeV for $\Xi^{--}\rightarrow \Sigma^-K^-$.
Similarly, we would expect ${\bf g}_{\Theta N} \approx 0.25$ and ${\bf
g}_{K\Theta N}\approx 2.35$, and we therefore predict a very 
{\it narrow} width of 2.60 MeV for the decay $\Theta^+\rightarrow K^+n$.

The narrowness of the partial widths in the antidecuplet follows from
a generically small transition axial-charge of about $1/4$, resulting
into a $\pi$-$PN$ decay constant of about 3 in the antidecuplet. The 
smallness of the axial-charge follows from the small overlap between
the three and five quark states.

\vskip 0.5cm

{\bf Summary and Discussion\,\,.}
We started by emphasizing that instanton-induced t'Hooft interaction
imply diquark substructure of multiquark hadrons and dense hadronic
matter,
with marked preference to the lowest flavor representations possible.
We then summarized the finding of ref.~\cite{RILM:BaryCor}:
in the instanton liquid model whereby there are two kinds of
deeply bound diquarks, the scalar and the (less bound) tensor.

We have then developed a schematic additive model, whereby diquarks
appear as building blocks, on equal footing with constituent quarks.
In such a model pentaquarks  are treated as 3-body states, so that their
classification in color and flavor becomes analogous to that of the
 baryons. If one uses two scalar diquarks, as 
  suggested by Jaffe and Wilczek, 
the P-wave is inevitable which seems to produce
states heavier than the ones reported, even in a  simple additive model
with very light diquarks. If one uses one
scalar and one tensor diquarks, the masses look more reasonable.
However, then the ensuing flavor representations are large, and although 
recently discovered quartet of $\Xi(1860)$ fits very well into this
model, the $\Theta^+$ has (so far) unobserved partners. 

We have related the widths  with the ``axial overlap''charge,
and  have argued that current data restrict it to be significantly
smaller than the nucleon axial charge, by about a factor of 3. This 
means that the Skyrme-model interpretation of pentaquarks, as a
Goldstone boson moving on top of the baryon is inadequate.

If one goes a step further, to 6-quark states,
for example by combining the proton and the neutron, one gets 3 $ud$ diquarks.
Again the asymmetric color wave function asks for another
asymmetry: to do so one can put all 3 diquarks into the P-wave state,
with the spatial wave function $\epsilon_{ijk}\partial_i \subs \partial_j \subs \partial_k
\subs$  suggested in the second paper of \cite{DIQ:BoseCond}.
This will cost $3(\Sigma+\delta M_{L=1})=2700\, MeV$, 
well in agreement with the magnitude of the
repulsive nucleon-nucleon core.
However if one considers the quantum
numbers of the famous $H$ dibaryon, one can also make those out of diquarks
such as $\subs \subd \subu$. The resulting wave function is overall
flavor antisymmetric with all diquarks in S-states. Thus there is no
need for P-wave or tensor diquarks for the $H$ dibaryon. Our schematic model would then
lead to a very light $H$, in contradiction to both experimental limits and
lattice results.

This last observation  calls for the lesson with which we would like to 
conclude our paper: all schematic models (including our own)
assume additivity of the constituents.
However, as we emphasized in Fig.1, due to the Pauli exclusion principle
 one instanton can only make one deeply bound 
diquark at a time. 
Thus, there must be  a {\em diquark-diquark repulsive core}.
One particular 3-body instanton repulsion effect was already discussed for 
the $H$ in \cite{Tak_Oka}. Multi-body instanton induced interactions
were also observed in heavy-light systems~\cite{heavy}. A generic way
to address these effects would be some dynamical
studies, directly antisymmetrizing 
5 or 6 quarks themselves, as well as with those  in the QCD vacuum
(unquenching). The evaluation of the pertinent correlators
on the lattice is badly needed: studies of
inter-diquark interactions in the  instanton
liquid model will be reported elsewhere~\cite{Pertot}.
Only with the resulting core potential included,
the diquark-based description of multiquark states and of dense 
quark matter may become truly quantitative.

\section{Acknowledgments}
We thank D.~Pertot for multiple valuable discussions and S.~Sasaki
for a helful correpondence.
This work was partially supported by the US DOE grant DE-FG-88ER40388.

\end{document}